\documentclass{amsart}
\usepackage{mathrsfs}
\usepackage{amssymb}
\usepackage{amsmath}
\usepackage{amsfonts}
\usepackage{enumerate}
\usepackage{pifont}
\usepackage{enumerate}
\usepackage[pdftex]{graphicx}
\usepackage{verbatim}
\usepackage{amsthm}
\usepackage[all]{xy}

\numberwithin{equation}{section}

\newtheorem{definition}{Definition}[section]

\newtheorem{proposition}{Proposition}[section]
\newtheorem{remark}{Remark}[section]
\newtheorem{example}{Example}[section]

\newcommand{\8}{\infty}
\newcommand{\el}{\ell}

\newcommand{\be}{\begin{eqnarray*}}
\newcommand{\ee}{\end{eqnarray*}}
\newcommand{\beq}{\begin{equation}}
\newcommand{\eeq}{\end{equation}}
\newcommand{\beqn}{\begin{equation*}}
\newcommand{\eeqn}{\end{equation*}}
\newcommand{\bs}{\begin{split}}
\newcommand{\es}{\end{split}}

\begin{document}

\title{Non-Abelian observable-geometric phases and\\ the Riemann zeros}

\author{Zeqian Chen}

\address{Wuhan Institute of Physics and Mathematics, Innovation Academy for Precision Measurement Science and Technology, Chinese Academy of Sciences, 30 West District, Xiao-Hong-Shan, Wuhan 430071, China.}

\thanks{Key words: The Riemann zeros; the Hilbert-P\'{o}lya conjecture; Floquet quantum system; observable-geometric phase; non-Abelian quantum connection.}

\date{}
\maketitle
\markboth{Zeqian Chen}%
{Quantum phase}

\begin{abstract}
The Hilbert-P\'{o}lya conjecture asserts that the imaginary parts of the nontrivial zeros of the Riemann zeta function (the Riemann zeros) are the eigenvalues of a self-adjoint operator (a quantum mechanical Hamiltonian, in the physical sense), as a promising approach to prove the Riemann hypothesis (cf.\cite{SH2011}). Instead of the eigenvalues, in this paper we consider observable-geometric phases as the realization of the Riemann zeros in a periodically driven quantum system, which were introduced in \cite{Chen2020} for the study of geometric quantum computation. To this end, we further introduce the notion of non-Abelian observable-geometric phases, involving which we give an approach to finding a physical system to study the Riemann zeros. Since the observable-geometric phases are connected with the geometry of the observable space according to the evolution of the Heisenberg equation, this sheds some light on the investigation of the Riemann hypothesis.
\end{abstract}

\maketitle


\section{Introduction}\label{Intro}

Let $\pi (x)$ denote the number of primes less than $x$ for every $x >0.$ Since 18th century, the asymptotic behavior of $\pi (x)$ as $x \to \8$ was one of the most important problem in mathematics. Gauss had conjectured that
\beq\label{eq:PNT}
\pi (x) \sim \frac{x}{\log x},\quad x \to \8,
\eeq
that is, $\lim_{x \to \8} \pi(x) / \frac{x}{\log x} =1.$ In 1859, Riemann gave an outline of a possible proof of Gauss's conjecture \eqref{eq:PNT}, utilizing the analytic function $\zeta$ of complex variables $s \in \mathbb{C},$ defined in the domain $\mathfrak{R} (s)>1$ ($\mathfrak{R} (s)$ denotes the real part of a complex number $s$) by the series
\beq\label{eq:Zeta}
\zeta (s) = \sum_{n=1}^\8 \frac{1}{n^s},
\eeq
and in $\mathbb{C}\setminus \{s=1\}$ by analytic continuation, now known as the Riemann zeta function. To this end, he made a conjecture that except for the trivial zeros $s=-2k$ ($k$'s are positive integers), the nontrivial zeros of the Riemann zeta function $\zeta$ (the Riemann zeros) are at the line $\mathfrak{R} (s) = \frac{1}{2}.$ This was the Riemann hypothesis. Based on Riemann's work, Hadamard and de la Vall\'{e}e Poussin proved independently Gauss's conjecture \eqref{eq:PNT} in 1896, through showing that there is no Riemann zero at the line $\mathfrak{R}(s) =1.$ This famous result is called the prime number theorem. Further, in 1902, von Koch showed that the Riemann hypothesis implies
\beq\label{eq:RHPNT}
\pi (x) \sim \frac{x}{\log x} + C x^\frac{1}{2} \log x, \quad x \to \8,
\eeq
where $C>0$ is an absolute constant, and vice verse. Therefore, the Riemann hypothesis yields a much better error term \eqref{eq:RHPNT} than the prime number theorem \eqref{eq:PNT}. We refer to \cite{Bro2017} for the details.

Concerning the Riemann hypothesis, Hilbert and P\'{o}lya suggested independently that the Riemann zeros correspond to the eigenvalues of a self-adjoint operator (or a quantum mechanical Hamiltonian, in the physical sense) around 1910 (cf.\cite[p.2]{ES2010}). This was called the Hilbert-P\'{o}lya conjecture, as a promising approach to prove the Riemann hypothesis that has been tried repeatedly (cf.\cite{Sierra2019} and references therein). Roughly speaking, Connes \cite{Connes1999}, Berry and Keating \cite{BK1999a, BK1999b} initiated the investigation of the Hilbert-P\'{o}lya conjecture in 1999, involving the Connes-Berry-Keating operator (the so-called Berry-Keating Hamiltonian, in physical literatures) as follows
\beq\label{eq:CBK}
H_\mathrm{CBK} = - \mathrm{i} \Big (x \frac{d}{d x} + \frac{1}{2} \Big )
\eeq
in the Hilbert space $L^2 (\mathbb{R})$ (see \cite{CC2021} for more details). Then, Bender, Brody, and M\"{u}ller \cite{BBM2017} in 2017 introduced a (PT-symmetric) Hamiltonian that is a similarity transformation of $H_\mathrm{CBK}$ restricted in $L^2[0,\8),$ such that its eigenvalues formally realize the Riemann zeros under some physical conditions. Recently, Yakaboylu \cite{Yak2024} introduced another Hamiltonian that is also a similarity transformation of $H_\mathrm{CBK}$ in $L^2 [0, \8),$ such that its eigenvalues can be involved as the realization of the Riemann zeros. However, the self-adjointness of both Hamiltonians remains open.

Instead of the eigenvalues, in this paper we consider observable-geometric phases as the realization of the Riemann zeros in a periodically driven quantum system, namely a Floquet quantum system. The observable-geometric phases are a sequence of geometric phases for the observable in a quantum system, which were introduced in \cite{Chen2020} for the study of geometric quantum computation (cf.\cite{SMC2016}). To this end, we further introduce the notion of non-Abelian observable-geometric phases, which is defined as a sequence of unitary operators associated with a complete set of eigenspaces of the observable. Using the non-Abelian observable-geometric phases, we will give an approach to finding a physical system to study the Riemann zeros. In contrast to that classical/Riemannian geometry of the state space \cite{Simon1983, BMKNZ2003} is suited to the usual geometric phases for the state \cite{Berry1984, AA1987}, the geometry suited for the observable-geometric phases is a kind of quantum/operator geometric structure over the observable space according to the evolution of the Heisenberg equation \cite{Chen2021, CSY2023}.  This should shed some light on the investigation of the Riemann hypothesis.

The paper is organized as follows. In Section \ref{NOGP}, we will introduce the definition of non-Abelian observable-geometric phases and discuss its elementary properties. Sections \ref{Ex} will present an example for non-Abelian observable-geometric phases in a three-level Floquet system. In Section \ref{Ro}, we will give an approach to finding a quantum system to study the Riemann zeros via the non-Abelian observable-geometric phases. We will give a summary in Section \ref{Concl}. Finally, we will include an appendix, namely Section \ref{App}, on the geometry of non-Abelian operator-principal fiber bundles, which is needed for a geometrical description of the non-Abelian observable-geometric phase.

\section{Non-Abelian observable-geometric phase}\label{NOGP}

For the sake of simplicity, we only consider finite quantum systems, namely the associated Hilbert spaces $\mathbb{H}$ have a finite dimension. In what follows, we always denote by $\mathcal{B} (\mathbb{H})$ the algebra of all bounded operators on $\mathbb{H},$ by $\mathcal{O} (\mathbb{H})$ the set of all self-adjoint operators on $\mathbb{H},$ and by $\mathcal{U} (\mathbb{H})$ the group of all unitary operators on $\mathbb{H}.$ Without specified otherwise, the integer $d$ always denotes the dimension of $\mathbb{H},$ and $I$ the identity operator on $\mathbb{H}.$

Consider a finite quantum system with a time-dependent Hamiltonian $H(t).$ Since the associated Hilbert space $\mathbb{H}$ has finite dimension $d,$ each Hamiltonian $H(t)$ has a discrete spectrum as well as all observables considered in the following. Indeed, the spectrum of $H(t)$ at any given time will not be of importance. Instead, we shall consider the time evolution operator, as called propagator, generated by $H(t)$ (see \cite[Theorem X.69]{RS1980II}). This is a two-parameter continuous family $\{U(t,s): t,s \in \mathbb{R} \}$ of unitary operators such that for any $t, r, s \in \mathbb{R},$
\begin{equation}\label{eq:Propagator}
U(t,t) = I,\quad U(t, r) U (r, s) = U(t,s),
\end{equation}
and satisfying the operator Schr\"{o}dinger equation
\begin{equation}\label{eq:SchrodingerEquPropagator}
\mathrm{i} \frac{d}{d t} U(t,s) = H(t) U(t,s).
\end{equation}
Then for any $s \in \mathbb{R}$ and $\phi \in \mathbb{H},$ $\phi_s (t) = U (t,s) \phi$ is the unique solution of the time-dependent Schr\"{o}dinger equation
\begin{equation}\label{eq:SchrodingerEquTime}
\mathrm{i} \frac{d}{d t} \phi_s(t) = H(t) \phi_s (t)
\end{equation}
with $\phi_s (s) = \phi.$

Given any observable $X_0,$ namely a self-adjoint operator on $\mathbb{H},$ by \eqref{eq:SchrodingerEquPropagator} we conclude that $X(t) = U(0,t) X_0 U(t,0)$ is the unique solution of the time-dependent Heisenberg equation
\begin{equation}\label{eq:HeisenbergEquTime}
\mathrm{i} \frac{d X (t)}{d t} = [X (t), \tilde{H}(t)]
\end{equation}
with $X(0) = X_0,$ where $\tilde{H}(t) = U(0,t) H(t) U(t, 0).$ If there exists $T>0$ such that $X(T) = X(0),$ the time evolution of observable $(X(t): t \in \mathbb{R})$ is then called cyclic with period $T,$ and $X_0 = X(0)$ is said to be a cyclic observable.

Let $X_0$ be an observable with the spectral decomposition $X_0 = \sum^n_{j=1} \lambda_j E_j$, where $\lambda_j$'s are different (degenerate or non-degenerate) egenvalues of $X_0$ and $E_j$'s are the associated spectral projections. Assume that $X(t) = U(0,t) X_0 U(t,0)$ is cyclic with period $T,$ namely $X(T) = X_0$ so that
\beq\label{eq:SpectProj}
U(0,T) E_j U(T,0)=E_j,\quad 1\le j \le n.
\eeq
Denote $d_j$ to be the dimension of $E_j [\mathbb{H}]$ for $1\le j \le n$. Then $\sum_j d_j = d$. For $0\le t \le T,$ let $\mathbb{H}_j (t) = U(0,t)E_j U(t,0) [\mathbb{H}]$ of the dimension $d_j$ for $1 \le j \le n$. By \eqref{eq:SpectProj} one has
\be
\mathbb{H}_j(T) = \mathbb{H}_j (0) = E_j[\mathbb{H}], \quad 1 \le j \le n.
\ee
Note that $\mathbb{H} = \bigoplus_j \mathbb{H}_j (t)$ for all $0 \le t \le T.$

For $1 \le j \le n$, let $\{\psi^{(j)}_k: k=1,\ldots, d_j\}$ be an orthonormal basis of $E_j [\mathbb{H}]$. Denoting $\psi^{(j)}_k (t) = U(0,t) \psi^{(j)}_k$ for $1 \le k \le d_j,$ which are the eigenstates of $X(t)$ associated with the eigenvalue $\lambda_j$, we conclude that $\psi^{(j)}_k (t)$ satisfies the skew (time-dependent) Schr\"{o}dinger equation
\begin{equation}\label{eq:SchrodingerEquTimeEigenstate}
\mathrm{i} \frac{d}{d t} \psi^{(j)}_k (t) = - \tilde{H}(t) \psi^{(j)}_k (t),\quad 1 \le k \le d_j,
\end{equation}
with $\psi^{(j)}_k (0) = \psi^{(j)}_k,$ due to the fact that $U(0,t) = U(t,0)^{-1}.$ By \eqref{eq:SpectProj}, one has that for $0 \le t \le T,$ $\{\psi^{(j)}_k (t): k=1,\ldots, d_j\}$ is an orthonormal basis of $\mathbb{H}_j (t)$ for $1\le j \le n.$ In particular, $\{\psi^{(j)}_k (T): k=1,\ldots, d_j\}$ is an orthonormal basis of $\mathbb{H}_j (0)$ ($j=1, \ldots,n$).

For $1 \le j \le n,$ let $\tilde{V}_j(t) = [\tilde{v}^{(j)}_{m k} (t)]^{d_j}_{m,k=1}$ be the solution to the matrix equation
\beq\label{eq:GeoPhaseMatrix}
\mathrm{i} \frac{d}{d t} \tilde{V}_j (t) = C_j (t) \tilde{V}_j (t),
\eeq
with $\tilde{V}_j(0) = I_{n_i}$, where $C_j (t) = [c^{(j)}_{m k} (t)]^{d_j}_{m,k=1}$ with
\be
c^{(j)}_{m,k} (t) = \langle \psi^{(j)}_m (t), \tilde{H} (t) \psi^{(j)}_k (t)\rangle = \langle \psi^{(j)}_m, H(t) \psi^{(j)}_k \rangle, \quad 1 \le m,k \le d_j.
\ee
All $\tilde{V}_j (t)$'s are $d_j \times d_j$ unitary matrices, since $C_j (t)$'s are all Hermitian matrices. Define
\beq\label{eq:ParallCurveState}
\tilde{\psi}^{(j)}_k (t) = \sum^{d_j}_{m=1} \tilde{v}^{(j)}_{m k} (t) \psi^{(j)}_m (t),\quad 1 \le k \le d_j.
\eeq
For each $0 \le t \le T$, $\{\tilde{\psi}^{(j)}_k (t): k =1, \ldots, d_j\}$ is an orthonormal basis of $\mathbb{H}_j (t)$ for $1\le j \le n$.

\begin{proposition}\label{prop:ParallCondState}\rm
With the above notations, for $1 \le j \le n$,
\beq\label{eq:ParallCondState}
\langle \tilde{\psi}^{(j)}_m (t),\frac{d}{d t} \tilde{\psi}^{(j)}_k (t)\rangle = 0,\quad \forall 1\le m,k \le d_j.
\eeq
This shows that $\tilde{\psi}^{(j)}_m (t)$'s are ``the parallel transportation" in some sense (see Proposition \ref{prop:GeoInterNaOGP} below).
\end{proposition}

\begin{proof}
By \eqref{eq:SchrodingerEquTimeEigenstate}, one has
\be\begin{split}
\frac{d}{d t} \tilde{\psi}^{(j)}_k (t) =\sum^{d_j}_{m=1} \Big [ \Big ( - \mathrm{i} \sum^{d_j}_{\el =1} \langle \psi^{(j)}_m (t), \tilde{H} (t) \psi^{(j)}_\el (t)\rangle \tilde{v}^{(j)}_{\el k} (t)\Big ) \psi^{(j)}_m (t) + \mathrm{i} \tilde{v}^{(j)}_{m k} (t) \tilde{H} (t) \psi^{(j)}_m (t) \Big ].
\end{split}\ee
Taking the inner product of both sides of the above equation with $\psi^{(j)}_m (t)$, we obtain
\beq\label{eq:ParallCondStateEvo}
\langle \psi^{(j)}_m (t),\frac{d}{d t} \tilde{\psi}^{(j)}_k (t)\rangle = 0,\quad \forall 1\le m,k \le d_j.
\eeq
Thus, by \eqref{eq:ParallCurveState} we obtain \eqref{eq:ParallCondState}.
\end{proof}

Since $\{\psi^{(j)}_k (T): k=1,\ldots, d_j\}$ is an orthonomal basis of $\mathbb{H}_j (0),$ so does $\{\tilde{\psi}^{(j)}_k (T): k=1,\ldots, d_j\}$, we have
\be
\tilde{\psi}^{(j)}_k (T) = \sum^{d_j}_{m=1} g^{(j)}_{m k}\psi^{(j)}_m,
\ee
where $g^{(j)}_{m k}= \langle \psi^{(j)}_m, \tilde{\psi}^{(j)}_k (T) \rangle$ for all $1 \le m,k \le d_j.$ This leads to the notion of non-Abelian geometric phase for the observable as follows.

\begin{definition}\label{df:nAOGP}
With the above notations, the non-Abelian geometric phases of the periodic evolution of observable $X(t)$ is defined by
\begin{equation}\label{eq:nAOGP}
G_j = [g^{(j)}_{m k}]^{d_j}_{m,k =1},
\end{equation}
for every $j = 1, \ldots, n.$ We simply call $G_j$'s the non-Abelian observable-geometric phases (NOGPs, in short).
\end{definition}

The following proposition shows that $G_j$ is geometrical in the physical sense that it is independent of the Hamiltonian $H(t)$ and depends only on the Hilbert subspace loop $\mathbf{K}_j: [0,T]\ni t \mapsto \mathbb{H}_j(t)$.

\begin{proposition}\label{prop:nAOGP}\rm
With the above notations, for $1 \le j \le n$, $G_j$ is a $d_j \times d_j$ unitary matrix and depends only on the loop $\mathbf{K}_j: [0,T]\ni t \mapsto \mathbb{H}_j(t)$. Precisely,
\beq\label{eq:nAOGPclosedRep}
G_j = I_{d_j} + \sum_{m=1}^\8 \mathrm{i}^m \int^T_0 \int^{t_1}_0 \cdots \int^{t_{m-1}}_0 A_j (t_1) \cdots A_j (t_m) d t_m \cdots d t_1,
\eeq
where $A_j (t) = [a^{(j)}_{m k} (t)]^{d_j}_{m,k=1}$ with $a^{(j)}_{m k} (t) = \mathrm{i} \langle \bar{\psi}^{(j)}_m (t), \frac{d}{d t} \bar{\psi}^{(j)}_k (t)\rangle$, and $\{\bar{\psi}^{(j)}_k (t): k=1,\ldots, d_j\}$ is an orthonormal basis of $\mathbb{H}_j (t)$ such that for each $1\le k \le d_j,$ the mapping $[0,T] \ni t \mapsto \bar{\psi}^{(j)}_k (t)$ is continuously differential $\mathbb{H}$-valued function satisfying $\bar{\psi}^{(j)}_k (T)=\bar{\psi}^{(j)}_k (0) = \psi^{(j)}_k.$
\end{proposition}

\begin{proof}
Since $\{\tilde{\psi}^{(j)}_k (T): k=1,\ldots, d_j\}$ and $\{\psi^{(j)}_k: k=1,\ldots, d_j\}$ are both orthonomal bases of $\mathbb{H}_j (0),$ it follows that $G_j$ is a $d_j \times d_j$ unitary matrix. Since $\{\psi^{(j)}_k (t): k=1,\ldots, d_j\}$ is an orthonormal basis of $\mathbb{H}_j (t)$, by \eqref{eq:ParallCondStateEvo} one has
\beq\label{eq:ParallCondStateClosed}
\langle \bar{\psi}^{(j)}_m (t),\frac{d}{d t} \tilde{\psi}^{(j)}_k (t)\rangle = 0,
\eeq
for any $1 \le m,k \le d_j$.

Next, since $\{\tilde{\psi}^{(j)}_k (t): k=1,\ldots, d_j\}$ is also an orthonormal basis of $\mathbb{H}_j (t)$, we obtain
\beq\label{eq:ParaClosedState}
\tilde{\psi}^{(j)}_k (t) = \sum^{d_j}_{p=1} \bar{v}^{(j)}_{p k} (t) \bar{\psi}^{(j)}_p (t)
\eeq
where $\bar{v}^{(j)}_{mk}(t) = \langle \bar{\psi}^{(j)}_m (t), \tilde{\psi}^{(j)}_k (t) \rangle$. By \eqref{eq:ParaClosedState}, one has
\be
\frac{d}{d t} \tilde{\psi}^{(j)}_k (t) =\sum^{d_j}_{p=1} \Big ( \Big [ \frac{d}{d t}\bar{v}^{(j)}_{p k} (t) \Big ] \bar{\psi}^{(j)}_p (t) + \bar{v}^{(j)}_{p k} (t) \frac{d}{d t}\bar{\psi}^{(j)}_p (t) \Big ).
\ee
Taking the inner product of both sides of the above equation with $\bar{\psi}^{(j)}_m (t)$, by \eqref{eq:ParallCondStateClosed} we obtain
\be\begin{split}
0 = \frac{d}{d t} \bar{v}^{(j)}_{m k} (t) - \mathrm{i} \sum^{d_j}_{p=1} a^{(j)}_{m p} \bar{v}^{(j)}_{p k} (t), \quad \forall 1 \le m,k \le d_j.
\end{split}\ee
Thus,  let $\bar{V}_j (t) = [\bar{v}^{(j)}_{mk}(t)]^{d_j}_{m, k=1}$, we obtain
\beq\label{eq:ParaMatrixEq}
\mathrm{i}\frac{d}{d t} \bar{V}_j (t) = - A_j (t) \bar{V}_j (t).
\eeq
By the Dyson expansion (cf. \cite[Theorem X.69]{RS1980II}), we have
\be
\bar{V}_j (t) = I_{d_j} + \sum_{m=1}^\8 \mathrm{i}^m \int^t_0 \cdots \int^{t_{m-1}}_0 A_j (t_1) \cdots A_j (t_m) d t_m \cdots d t_1,\quad \forall 0 \le t \le T,
\ee
with the convention $t_0 = t$.

Since
\be
\tilde{\psi}^{(j)}_k (T) = \sum^{d_j}_{m=1} \bar{v}^{(j)}_{m k} (T) \psi^{(j)}_m = \sum^{d_j}_{m=1} g^{(j)}_{m k} \psi^{(j)}_m,\quad \forall 1 \le k \le d_j,
\ee
we conclude $G_j = \bar{V}_j (T)$, i.e., \eqref{eq:nAOGPclosedRep} holds true. Note that $G_j$ is independent of $\{\bar{\psi}^{(j)}_k (t): 0 < t \le T; k=1,\ldots, d_j \}$, while $\bar{V}_j (T)$  is independent of the Hamiltonian $H(t)$. But the choice of $\{\bar{\psi}^{(j)}_k (t): 0 \le t \le T; k=1,\ldots, d_j \}$ depends on the Hilbert subspace loop $\mathbf{K}_j: [0,T]\ni t \mapsto \mathbb{H}_j(t)$. Hence, $G_j$ is independent of the Hamiltonian $H(t)$ and depends only on the loop $\mathbf{K}_j: [0,T]\ni t \mapsto \mathbb{H}_j(t)$.
\end{proof}

\begin{remark}\rm
\begin{enumerate}[{\rm 1)}]

\item If one eigenvalue $\lambda_j$ of the initial observable $X_0$ is non-degenerate as the eigenstate $\psi^{(j)}$, the corresponding observable-geometric phase $G_j =[g^{(j)}]$ is a number such that $g^{(j)} = e^{\mathrm{i}\beta_j}$, where $\beta_j$ coincides with the observable-geometric phase defined in \cite{Chen2020} associated with the eigenstate $\psi^{(j)}$ of $X_0.$

\item For any $1 \le j \le n,$ we define
\beq\label{eq:jthHoloUop}
\tilde{U}_j = \sum^{d_j}_{m,k=1} g^{(j)}_{m k} |\psi^{(j)}_m\rangle \langle \psi^{(j)}_k|,
\eeq
then $\tilde{U}_j$ is a unitary operator on $\mathbb{H}_j (0)$. Thus,
\be
\tilde{U} = \sum^n_{j=1} \tilde{U}_j
\ee
is a unitary operator on $\mathbb{H}.$ In fact, $\tilde{U}$ is a {\it holonomy element} of a non-Abelian quantum connection in an operator-principal fiber bundle associated with the initial basis $\{\psi^{(j)}_k: 1\le k \le d_j; 1\le j \le n\}$ (see Section \ref{App} for the details).
\end{enumerate}
\end{remark}

If the system evolves adiabatically (cf.\cite[Chapter 2]{BMKNZ2003}), $H(t)$ varies slowly with $H(t) |\psi_n(t)\rangle= \lambda_n (t) |\psi_n(t)\rangle,$ for a complete set $\{|\psi_n(t)\rangle \},$ such that the state remains an eigenstate of $H(t)$ at all time $t$ with the same energy quantum number $n,$ namely the time evolution operator
\beq\label{eq:AdiabaticEvoOp}
U(t,0) \backsimeq \sum_n |\psi_n(t)\rangle \langle \psi_n(0)|
\eeq
to a good approximation, see \cite[(5)-(6)]{AA1987} or \cite[(2.37)-(2.39)]{BMKNZ2003} for the details. Suppose that $H(0)$ has a spectral decomposition $H(0) = \sum^n_{j=1} \lambda_j E_j$, where $\lambda_j$'s are different (degenerate or non-degenerate) egenvalues of $H(0)$ and $E_j$'s are the associated spectral projections. If the adiabatic evolution is cyclic with period $T,$ namely $H(0) = H(T),$ then $U(T,0)=I$ and so, the time-observable evolution $[0,T] \ni t \mapsto U(0,t) H(0) U(t,0)$ is cyclic with the period $T.$ In this case, we can obtain the non-Abelian observable-geometric phases of this observable evolution as constructed in Definition \ref{df:nAOGP}, but in general such phases do not coincide with those defined by Wilczek and Zee \cite{WZ1984}.

Our method of the above construction can apply to obtain the non-Abelian geometric phases for the state evolution introduced in both \cite{Anandan1988} and \cite{WZ1984}. To this end, let $U(t) = U(t,0)$ be the time evolution operator associated with the Hamiltonian $H(t).$ Let $\mathbb{H}(0)$ be the initial state space of the system with the initial basis $\{\psi_j: 1 \le j \le n\}$ of $\mathbb{H}(0)$ such that the Hilbert subspace flow: $[0,T] \ni t \mapsto \mathbb{H} (t) = \{U(t)x: x \in \mathbb{H}(0) \}$ is cyclic, namely $\mathbb{H}(T) = \mathbb{H}(0).$ Let $\tilde{V} (t) = [\tilde{v}_{m k} (t)]^n_{m,k=1}$ be the solution to the matrix equation
\beq\label{eq:GeoPhaseMatrixAdiabatic}
\mathrm{i} \frac{d}{d t} \tilde{V} (t) = C (t) \tilde{V} (t),
\eeq
with $\tilde{V}(0) = I_n$, where $C (t) = [c_{m k} (t)]^n_{m,k=1}$ with
\be
c_{m,k} (t) = - \langle U(t) \psi_m, H(t) U(t) \psi_k \rangle,\quad 1 \le m,k \le n.
\ee
Define
\beq\label{eq:ParallCurveStateAdiabatic}
\tilde{\psi}_k (t) = \sum^n_{m=1} \tilde{v}_{m k} (t) \psi_m (t),\quad 1 \le k \le n,
\eeq
where $\psi_m (t) = U(t) \psi_m$ for $1 \le m \le n.$ Since $\{\tilde{\psi}_k (T): k =1, \ldots, n\}$ is an orthonormal basis of $\mathbb{H} (0)$, we have
\be
\tilde{\psi}_k (T) = \sum^n_{m=1} g_{m k}\psi_m,
\ee
where $g_{m k}= \langle \psi_m, \tilde{\psi}_k (T) \rangle$ for all $1 \le m,k \le n.$ The non-Abelian geometric phase of this state evolution is then defined to be the matrix as follows
\begin{equation}\label{eq:nAOGPadiabatic}
G = [g_{m k}]^n_{m,k =1},
\end{equation}
which coincides with the non-Abelian geometric phase as defined in \cite{WZ1984, Anandan1988}. By the argument in Proposition \ref{prop:nAOGP}, we can show that this $G$ is independent of $H(t)$'s and depends only on the Hilbert subspace loop $\mathbf{K}: [0,T] \ni t \mapsto \mathbb{H} (t).$

\section{An example}\label{Ex}

For illustrating our approach to the NOGPs, we consider a three-level system, i.e., the Hilbert space $\mathbb{H} = \mathbb{C}^3$ with the standard basis
\be
|0\rangle = \left ( \begin{matrix} 1 \\
0\\
0
\end{matrix}\right ),\; |1\rangle = \left ( \begin{matrix} 0\\
1\\
0
\end{matrix}\right ),\;|2\rangle = \left ( \begin{matrix} 0 \\
0\\
1
\end{matrix}\right ).
\ee
Suppose that the Hamiltonian describing the system is given by (cf.\cite{STAHJS2012})
\beq\label{eq:3-levelHamiltonian}
H(t) = \Omega (t) (|2\rangle\langle b| + |b \rangle\langle 2|),
\eeq
where $|b\rangle = \overline{\omega}_0 |0\rangle + \overline{\omega}_1 |1\rangle$ with $\omega_0$ and $\omega_1$ being two constants such that $|\omega_0|^2 + |\omega_1|^2 =1$, and $\Omega: t \mapsto \Omega (t) \in \mathbb{R}$ is a continuous real-valued $T$-periodic function such that $\int^T_0 \Omega (t) d t = \pi.$ The time evolution operator $U(t,0)$ is given by (cf.\cite{AS2022})
\beq\label{eq:1qubitEvoOp}
U(t,0) =|d \rangle \langle d | + \cos \Phi (t) (|b\rangle\langle b| + |2 \rangle\langle 2|) - \mathrm{i} \sin \Phi (t) (|2\rangle\langle b| + |b \rangle\langle 2|),
\eeq
where $|d\rangle = -\omega_1 |0\rangle + \omega_0 |1\rangle$ and $\Phi (t) = \int^t_0 \Omega (s) d s$. Let $X_0 = \lambda_1(|0\rangle\langle0| + |1\rangle\langle1|) + \lambda_2 |2\rangle\langle2|$ be an observable with $\lambda_1 \not = \lambda_2$. Then the observable evolution $[0,T] \ni t \mapsto X(t) = U(0,t)X_0 U(t,0)$ is cyclic with the period $T$, that is,
\be
U(0,T) E_j U(T,0) = E_j, \quad j=1,2
\ee
where $E_1 = |0\rangle\langle0| + |1\rangle\langle1|$ and $E_2 = |2\rangle\langle2|$.

For the initial basis
\begin{equation}\label{eq:InitalbasisQubit}
\left \{  \begin{split} \psi^{(1)}_1 = & \cos \frac{\phi}{2} |0\rangle + e^{\mathrm{i} \varphi}\sin \frac{\phi}{2} |1\rangle,\\
\psi^{(1)}_2 = & - e^{-\mathrm{i} \varphi} \sin \frac{\phi}{2} |0\rangle + \cos \frac{\phi}{2}|1\rangle,
\end{split}\right.
\end{equation}
in $\mathbb{H}_1 (0) = E_1 [\mathbb{C}^3],$ since
\be
c^{(1)}_{mk} = \langle \psi^{(1)}_m (t), \tilde{H} (t) \psi^{(1)}_k (t) \rangle = \langle \psi^{(1)}_m, H(t) \psi^{(1)}_k \rangle =0, \quad m,k =1,2,
\ee
by \eqref{eq:GeoPhaseMatrix} we have $\tilde{V}_1(t) = I_2$ and so
\be
\tilde{\psi}^{(1)}_k (t) = \psi^{(1)}_k (t) = U(0,t) \psi^{(1)}_k, \quad k=1,2,
\ee
for all $0 \le t \le T.$ Since $U(0,T)=|d \rangle \langle d | -|b\rangle\langle b| - |2 \rangle\langle 2|$, we obtain
\be\left \{  \begin{split}
\tilde{\psi}^{(1)}_1 (T) = & g^{(1)}_{1 1} \psi^{(1)}_1 + g^{(1)}_{2 1}\psi^{(1)}_2, \\
\tilde{\psi}^{(1)}_2 (T) = & \overline{g^{(1)}_{2 1}}\psi^{(1)}_1 - g^{(1)}_{1 1} \psi^{(1)}_2,
\end{split}\right.\ee
where
\be
\begin{split}
g^{(1)}_{1 1} = & (|\omega_1|^2 - |\omega_0|^2) \cos \phi - (e^{-\mathrm{i} \varphi}\omega_0 \overline{\omega_1} + e^{\mathrm{i} \varphi}\overline{\omega_0} \omega_1)\sin \phi,\\
g^{(1)}_{2 1} = & e^{\mathrm{i} \varphi}(|\omega_0|^2 - |\omega_1|^2) \sin \phi - 2 \omega_0 \overline{\omega_1} \cos^2 \frac{\phi}{2} + 2 e^{2\mathrm{i} \varphi} \overline{\omega_0} \omega_1 \sin^2 \frac{\phi}{2}.\\
\end{split}
\ee
Thus, we have
\beq\label{eq:3level-1NOGP}
G_1 = \left [ \begin{matrix}
g^{(1)}_{1 1} & \overline{g^{(1)}_{2 1}}\\
g^{(1)}_{2 1} & - g^{(1)}_{1 1}
\end{matrix} \right ] = \mathbf{n} \cdot \bar{\sigma},
\eeq
where $\mathbf{n} = (\mathrm{Re}(g^{(1)}_{2 1}), \mathrm{Im} (g^{(1)}_{2 1}), g^{(1)}_{1 1})$, $\vec{\sigma}=(\sigma_x, \sigma_y, \sigma_z)$ is the standard Pauli matrices:
\beq\label{eq:PauliMat}
\sigma_x = \left ( \begin{matrix} 0 & 1 \\
1 & 0
\end{matrix}\right ),\; \sigma_y = \left ( \begin{matrix} 0 & - \mathrm{i} \\
\mathrm{i} & 0
\end{matrix}\right ),\;\sigma_z = \left ( \begin{matrix} 1 & 0 \\
0 & -1
\end{matrix}\right ).
\eeq
Taking $\phi =0$, $\omega_0 = e^{\mathrm{i}\vartheta} \sin \frac{\theta}{2}$ and $\omega_1 = - \cos \frac{\theta}{2}$, one has
\beq\label{eq:3level1-gate}
G_1 = \left [ \begin{matrix}
\cos \theta & e^{-\mathrm{i}\vartheta} \sin \theta\\
e^{\mathrm{i}\vartheta} \sin \theta & -\cos \theta
\end{matrix} \right ] = \mathbf{n} \cdot \vec{\sigma}
\eeq
where $\mathbf{n} = (\sin \theta \cos \vartheta, \sin \theta \sin \vartheta, \cos \theta)$.

For the initial basis $\psi^{(2)}_1= e^{\mathrm{i}\phi} |2\rangle$ in $\mathbb{H}_2 (0) = E_2 [\mathbb{H}],$ since $\tilde{V}_2 (t)= 1$, we have $\tilde{\psi}^{(2)}_1 (t) = \psi^{(2)}_1 (t)$ and so
\be
\tilde{\psi}^{(2)}_1 (T) = - \psi^{(2)}_1.
\ee
Thus $G_2 = e^{\mathrm{i} \pi}$, where $\pi$ is equal to the observable-geometric phase defined in \cite{Chen2020} associated with the non-degenerate eigenstate $|2\rangle$ of the observable $X_0.$

\section{The Riemann zeros}\label{Ro}

In this section, we shall show that the NOGPs can be used to find the Riemann zeros. To this end, we consider the Riemann $\Xi$ function
\be
\xi (E) = \frac{1}{2} s (s-1) \pi^{-s/2} \Gamma (s/2) \zeta (s),\quad s= \frac{1}{2} + \mathrm{i} E,
\ee
whose zeros coincide with the Riemann zeros.  This can be written as
\beq\label{eq:Xi}
\xi (E) = \int^\8_0 \Phi (t) \cos (E t /2) d t,
\eeq
where
\be
\Phi (t) = 2 \pi e^{5 t /4} \sum^\8_{n=1} (2 \pi e^t n^2 - 3)n^2 e^{- \pi n^2 e^t}
\ee
(see \cite{Bro2017} for the details).

As in Section \ref{Ex}, we consider a three-level system associated with the Hilbert space $\mathbb{H} = \mathbb{C}^3.$ The Hamiltonian describing the system takes the form
\beq\label{eq:3LH}
H_E (t) = \Delta_E ( |0\rangle \langle 1| + |1 \rangle \langle 0|) + \Omega (t) (|2\rangle\langle b| + |b \rangle\langle 2|),
\eeq
where $\Delta_E = \xi (E)$ for any real parameter $E,$ $|b\rangle = \overline{\omega}_0 |0\rangle + \overline{\omega}_1 |1\rangle$ with $\omega_0$ and $\omega_1$ being two constants such that $|\omega_0|^2 + |\omega_1|^2 =1$, and $\Omega: t \mapsto \Omega (t) \in \mathbb{R}$ is a continuous real-valued $T$-periodic function such that $\int^T_0 \Omega (t) d t = \pi.$

When $s = \frac{1}{2} + \mathrm{i} E$ is a Riemann zero, $\Delta_E =0$ and then the Hamiltonian $H_E(t)$ reduces to
\be
H(t) = \Omega (t) (|2\rangle\langle b| + |b \rangle\langle 2|).
\ee
As shown in Section \ref{Ex}, in this case, we obtain the NOGPs as the same as the non-Abelian geometric phases given by Sj\"{o}qvist {\it et al} \cite{STAHJS2012}. Indeed, let $\omega_0 = e^{\mathrm{i}\vartheta} \sin \frac{\theta}{2}$ and $\omega_1 = - \cos \frac{\theta}{2}$, taking the initial basis $\psi^{(1)}_1 = |0\rangle$ and $\psi^{(1)}_2 = |1\rangle$ we obtain the non-Abelian observable-geometric phase $G_1$ in \eqref{eq:3level1-gate}, which defines the one-qubit gate as follows
\be
U^{(1)} (\mathbf{C}_\mathbf{n}) = \cos \theta |0 \rangle\langle 0| + e^{\mathrm{i} \vartheta} \sin \theta |1 \rangle\langle 0| + e^{-\mathrm{i} \vartheta} \sin \theta |0 \rangle\langle 1| - \cos \theta |1 \rangle\langle 1| = \mathbf{n} \cdot \vec{\sigma},
\ee
where $\mathbf{n} = (\sin \theta \cos \vartheta, \sin \theta \sin \vartheta, \cos \theta).$

Therefore, the Riemann zeros can be detected by realizing a non-Abelian observable-geometric phase or equivalently,  one-qubit geometric gate, as the real parameter $E$ varies in a three-level Floquet system. Note that, the first two Riemann zeros can be obtained by detecting the degeneracy of the quasienergies of a two-level Floquet system (cf.\cite{CS2015, HL2020, HL2021}). However, our model concerns the realization of the non-Abelian observable-geometric phases, that depend only on the geometrical properties of a Floquet quantum system.

\section{Conclusion}\label{Concl}

Geometric phases provide a new way of looking at quantum mechanics and have various applications in quantum information and related fields (cf.\cite{BMKNZ2003, PLC2022, Sj2015}). The usual theory of the geometric phase is based on the Schr\"{o}dinger picture, that is, the geometric phase is defined for the state and explained as the geometrical properties of the state space (cf.\cite{Berry1984, Simon1983, WZ1984, AA1987, SB1988, Anandan1988, Anandan1992}). In \cite{Chen2020}, we define the notion of the geometric phases for the observable (observable-geometric phase) in the Heisenberg picture, which is explained as the geometrical properties of the observable space. This provide an alternative way of studying the geometrical properties of the quantum system from the viewpoint of the observable. We have shown that the observable-geometric phases can be used to realize a universal set of quantum gates in quantum computation.

Here, we continue to study the so-called non-Abelian observable-geometric phases. A mathematical construction of the non-Abelian observable-geometric phases is presented. As application, we propose an approach to finding a three-level Floquet system to study the Riemann zeros via the non-Abelian observable-geometric phases. Following the Hilbert-P\'{o}lya conjecture, instead of the eigenvalues, we would conjecture that the Riemann zeros correspond to the observable-geometric phases of a (non-Hermitian) quantum evolution (cf.\cite{Chen2021}). Therefore, it may not be unreasonable to hope that this new insight of the observable-geometric phase may have heuristic value on the investigation of the Riemann hypothesis.

\section{Appendix: Geometry over non-Abelian quantum connection}\label{App}

In this appendix, we first introduce the notion of non-Abelian operator-principal fiber bundles. Then we define the notions of non-Abelian quantum connections and quantum parallel transportation. Finally, we give the geometrical description of non-Abelian observable-geometric phases in terms of the non-Abelian quantum connections.

\subsection{Operator-principal fiber bundles}\label{OpFB}

For an orthonomal basis $O_0 = \{e_n: 1\le n \le d\}$ of $\mathbb{H},$ a $[d_1, \ldots, d_n]$-partition of $Q_0$ consists of $n$ disjoint nonempty subsets $Q^{(1)}_0, \ldots, Q^{(n)}_0$ of $Q_0$, that is, $Q^{(j)}_0 = \{e^{(j)}_k: 1 \le k \le d_j \}$ ($j=1, \ldots,n$), $\sum^n_{j=1} d_j = d,$ $Q^{(j)}_0 \cap Q^{(k)}_0 = \emptyset$ if $j\not=k$, and $Q_0 = \cup^n_{j=1} Q^{(j)}_0$. Such a partition of $Q_0$ is denoted by $[Q^{(j)}_0]^n_{j=1}$ or $[Q^{(1)}_0, \ldots, Q^{(n)}_0].$

\begin{definition}\label{df:GaugeGroup}
Given an orthonomal basis $O_0$ of $\mathbb{H}$ with a fixed $[d_1, \ldots, d_n]$-partition $[Q^{(j)}_0]^n_{j=1}$ such that $Q^{(j)}_0 = \{e^{(j)}_k: 1 \le k \le d_j \}$ ($j=1, \ldots,n$), define
\be
\mathcal{G}^{[Q^{(j)}_0]^n_{j=1}}_{Q_0} = \Big \{ U \in \mathcal{U} (\mathbb{H}): U = \sum^n_{j=1}  \sum^{d_j}_{m,k=1} u^{(j)}_{m k} |e^{(j)}_m \rangle \langle e^{(j)}_k|,\; [u^{(j)}_{m k}]^{d_j}_{m,k=1} \in \mathbf{U}(d_j, \mathbb{C}), 1 \le j \le n \Big \},
\ee
where $\mathbf{U}(k, \mathbb{C})$ is the set of all $k \times k$ complex unitary matrices for an integer $k \ge 1$.
\end{definition}

Note that $\mathcal{G}^{[Q^{(j)}_0]^n_{j=1}}_{Q_0}$ is a topological subgroup of $\mathcal{U} (\mathbb{H})$, that is, each $U \in \mathcal{G}^{[Q^{(j)}_0]^n_{j=1}}_{Q_0}$ is of the form
\be
U = \sum^n_{j=1} U_j,
\ee
where $U_j \in \mathcal{U}(\mathbb{H}_j)$ with the Hilbert subspace $\mathbb{H}_j$ of $\mathbb{H}$ generated by $\{e^{(j)}_k: 1 \le k \le d_j \}$ for $1 \le j \le n.$

To construct an operator-principal fiber bundle, we fix an orthonomal basis $O_0$ of $\mathbb{H}$ with a given $[d_1, \ldots, d_n]$-partition $[Q^{(j)}_0]^n_{j=1}$. The right action of $\mathcal{G}^{[Q^{(j)}_0]^n_{j=1}}_{Q_0}$ on $\mathcal{U}(\mathbb{H})$ is defined as: For any $G \in \mathcal{G}^{[Q^{(j)}_0]^n_{j=1}}_{Q_0},$
$$
R_G: U \mapsto U G,\quad \forall U \in \mathcal{U} (\mathbb{H}).
$$
Since $R_{G_1 G_2} U = R_{G_2}(R_{G_1} U)$ and $R_I U = U$ for all $G_1, G_2 \in \mathcal{G}^{[Q^{(j)}_0]^n_{j=1}}_{Q_0}$ and $U \in \mathcal{U} (\mathbb{H}),$ then $\mathcal{U} (\mathbb{H})$ is a right $\mathcal{G}^{[Q^{(j)}_0]^n_{j=1}}_{Q_0}$-space.

Two unitary operator $U$ and $V$ in $\mathcal{U} (\mathbb{H})$ is called $\mathcal{G}^{[Q^{(j)}_0]^n_{j=1}}_{Q_0}$-equivalent provided there exists $G \in \mathcal{G}^{[Q^{(j)}_0]^n_{j=1}}_{Q_0}$ such that $U = V G$. This relation is an equivalent relation, and the set $U \mathcal{G}^{[Q^{(j)}_0]^n_{j=1}}_{Q_0} = \{U G : G \in \mathcal{G}^{[Q^{(j)}_0]^n_{j=1}}_{Q_0}\}$ is the equivalent class determined by $U \in \mathcal{U} (\mathbb{H}).$ We define
\be
\mathcal{W}^{[Q^{(j)}_0]^n_{j=1}}_{Q_0} = \{ U \mathcal{G}^{[Q^{(j)}_0]^n_{j=1}}_{Q_0}: U \in \mathcal{U} (\mathbb{H}) \}
\ee
with the quotient topology, that is, the largest topology such that the projection
\be
\Pi^{[Q^{(j)}_0]^n_{j=1}}_{Q_0}: \mathcal{U} (\mathbb{H}) \mapsto \mathcal{W}^{[Q^{(j)}_0]^n_{j=1}}_{Q_0}
\ee
is continuous. By definition (cf.\cite[Definition 4.1.6]{Huse1994}), the triple
\be
(\mathcal{U} (\mathbb{H}), \Pi^{[Q^{(j)}_0]^n_{j=1}}_{Q_0}, \mathcal{W}^{[Q^{(j)}_0]^n_{j=1}}_{Q_0})
\ee
is a $\mathcal{G}^{[Q^{(j)}_0]^n_{j=1}}_{Q_0}$-bundle, denoted by $\xi^{[Q^{(j)}_0]^n_{j=1}}_{Q_0}.$

\begin{proposition}\label{prop:OpFB}\rm
The $\mathcal{G}^{[Q^{(j)}_0]^n_{j=1}}_{Q_0}$-bundle $\xi^{[Q^{(j)}_0]^n_{j=1}}_{Q_0}$ is a principal $\mathcal{G}^{[Q^{(j)}_0]^n_{j=1}}_{Q_0}$-bundle.
\end{proposition}

\begin{proof}
Since the equality $U G = U$ for $G \in \mathcal{G}^{[Q^{(j)}_0]^n_{j=1}}_{Q_0}$ and $U \in \mathcal{U} (\mathbb{H})$ implies $G=I$, the right action of $\mathcal{G}^{[Q^{(j)}_0]^n_{j=1}}_{Q_0}$ on $\mathcal{U} (\mathbb{H})$ is free. Define $\tau: X^* = \{ (U, U G): U \in \mathcal{U} (\mathbb{H}), G \in \mathcal{G}^{[Q^{(j)}_0]^n_{j=1}}_{Q_0}\} \mapsto \mathcal{G}^{[Q^{(j)}_0]^n_{j=1}}_{Q_0}$ by $\tau (U, U G) = G$. Since the right action of $\mathcal{G}^{[Q^{(j)}_0]^n_{j=1}}_{Q_0}$ on $\mathcal{U} (\mathbb{H})$ is free, the mapping $\tau: X^* \mapsto \mathcal{G}^{[Q^{(j)}_0]^n_{j=1}}_{Q_0}$ is well defined, and then is a continuous translation function for the right $\mathcal{G}^{[Q^{(j)}_0]^n_{j=1}}_{Q_0}$-space $\mathcal{U} (\mathbb{H})$. Thus, by definition (cf.\cite[Definition 4.2.2]{Huse1994}), the $\mathcal{G}^{[Q^{(j)}_0]^n_{j=1}}_{Q_0}$-bundle $\xi^{[Q^{(j)}_0]^n_{j=1}}_{Q_0}$ is a principal $\mathcal{G}^{[Q^{(j)}_0]^n_{j=1}}_{Q_0}$-bundle.
\end{proof}

Note that the principal $\mathcal{G}^{[Q^{(j)}_0]^n_{j=1}}_{Q_0}$-bundle $\xi^{[Q^{(j)}_0]^n_{j=1}}_{Q_0}$ is a bundle with fiber $\mathcal{G}^{[Q^{(j)}_0]^n_{j=1}}_{Q_0}$, that is, for any $W \in \mathcal{W}^{[Q^{(j)}_0]^n_{j=1}}_{Q_0}$, $(\Pi^{[Q^{(j)}_0]^n_{j=1}}_{Q_0})^{-1} [W]$ is topologically homeomorphic to $\mathcal{G}^{[Q^{(j)}_0]^n_{j=1}}_{Q_0}$ (cf.\cite[Proposition 4.2.6]{Huse1994}). In this case, the group $\mathcal{G}^{[Q^{(j)}_0]^n_{j=1}}_{Q_0}$ is called the structure group of $\xi^{[Q^{(j)}_0]^n_{j=1}}_{Q_0}$ (in physical literatures, the structure group $\mathcal{G}^{[Q^{(j)}_0]^n_{j=1}}_{Q_0}$ is also called the gauge group of $\xi^{[Q^{(j)}_0]^n_{j=1}}_{Q_0}$, cf.\cite{BMKNZ2003}).

\begin{definition}\label{df:OpBF}
Given an orthonomal basis $O_0$ of $\mathbb{H}$ with a fixed $[d_1, \ldots, d_n]$-partition $[Q^{(j)}_0]^n_{j=1}$, the principal $\mathcal{G}^{[Q^{(j)}_0]^n_{j=1}}_{Q_0}$-bundle
\be
\xi^{[Q^{(j)}_0]^n_{j=1}}_{Q_0} = (\mathcal{U} (\mathbb{H}), \Pi^{[Q^{(j)}_0]^n_{j=1}}_{Q_0}, \mathcal{W}^{[Q^{(j)}_0]^n_{j=1}}_{Q_0})
\ee
is simply called an operator-principal fiber bundle (OPFB, in short) with the fiber $\mathcal{G}^{[Q^{(j)}_0]^n_{j=1}}_{Q_0}$.
\end{definition}

\begin{remark}\rm
Note that the operator-principal fiber bundle $\xi^{[Q^{(j)}_0]^n_{j=1}}_{Q_0}$ is not a principal fiber bundle in the usual sense of mathematical literatures (cf. \cite[Section 4.5]{Huse1994}).
\end{remark}

For two orthonomal bases $O_0= \{ e_n: n \ge 1 \}$ and $O'_0= \{ e_n': n \ge 1 \}$ of $\mathbb{H}$ with the same $[d_1, \ldots, d_n]$-partition $[Q^{(j)}_0]^n_{j=1}$ and $[Q'^{(j)}_0]^n_{j=1}$, we define a unitary operator $U_0$ by $U_0 e^{(j)}_k = e'^{(j)}_k$ for $1 \le k \le d_j$ and $j=1,\ldots, n.$ Then the map $(T,f): \xi^{[Q^{(j)}_0]^n_{j=1}}_{Q_0} \mapsto \xi^{[Q'^{(j)}_0]^n_{j=1}}_{Q'_0}$ defined by $T U = U_0 U U^{-1}_0$ and $f (U \mathcal{G}^{[Q^{(j)}_0]^n_{j=1}}_{Q_0}) = U_0 U U^{-1}_0 \mathcal{G}^{[Q'^{(j)}_0]^n_{j=1}}_{Q'_0}$ for $U \in \mathcal{U} (\mathbb{H})$ is a bundle isomorphism such that $T$ maps the fiber of $\xi^{[Q^{(j)}_0]^n_{j=1}}_{Q_0}$ over $U\mathcal{G}^{[Q^{(j)}_0]^n_{j=1}}_{Q_0}$ onto the fiber of $\xi^{[Q'^{(j)}_0]^n_{j=1}}_{Q'_0}$ over $U_0 U U^{-1}_0 \mathcal{G}^{[Q'^{(j)}_0]^n_{j=1}}_{Q'_0},$ namely the following diagram is commutative:
 \[
 \xymatrix{
\mathcal{U} (\mathbb{H}) \ar[d]_{\Pi^{[Q^{(j)}_0]^n_{j=1}}_{Q_0}} \ar[rr]^{T} & & \mathcal{U} (\mathbb{H}) \ar[d]^{\Pi^{[Q'^{(j)}_0]^n_{j=1}}_{Q'_0}}  \\
  \mathcal{W}^{[Q^{(j)}_0]^n_{j=1}}_{Q_0} \ar[rr]_{f}   & &     \mathcal{W}^{[Q'^{(j)}_0]^n_{j=1}}_{Q'_0}   .    }
 \]
Thus, $\xi^{[Q^{(j)}_0]^n_{j=1}}_{Q_0}$ and $\xi^{[Q'^{(j)}_0]^n_{j=1}}_{Q'_0}$ are isomorphic as principal bundles (cf.\cite{Isham1999}).

\subsection{Non-Abelian quantum connection}\label{nAqConnection}

As in \cite{Chen2020}, we need to define the suitable concepts of quantum connection and parallel transportation over the operator-principal fiber bundle $\xi^{[Q^{(j)}_0]^n_{j=1}}_{Q_0}$. In what follows, we will introduce a geometric structure over $\xi^{[Q^{(j)}_0]^n_{j=1}}_{Q_0}$ in a certain operator-theoretic sense, where the differential structure over $\mathcal{W}^{[Q^{(j)}_0]^n_{j=1}}_{Q_0}$ is different from the usual one (cf.\cite{Isham1999}).

At first, a tangent vector for $\mathcal{G}^{[Q^{(j)}_0]^n_{j=1}}_{Q_0}$ is defined in a operator-theoretic way, as in \cite[Definition A.2]{Chen2020}. The set of all tangent vectors of $\mathcal{G}^{[Q^{(j)}_0]^n_{j=1}}_{Q_0}$ at $U$ is denoted by $T_U \mathcal{G}^{[Q^{(j)}_0]^n_{j=1}}_{Q_0},$ and
\be
T \mathcal{G}^{[Q^{(j)}_0]^n_{j=1}}_{Q_0}= \bigcup_{U \in \mathcal{G}^{[Q^{(j)}_0]^n_{j=1}}_{Q_0}} T_U \mathcal{G}^{[Q^{(j)}_0]^n_{j=1}}_{Q_0}.
\ee
In particular, we denote $\mathrm{g}^{[Q^{(j)}_0]^n_{j=1}}_{Q_0} = T_U \mathcal{G}^{[Q^{(j)}_0]^n_{j=1}}_{Q_0}$ if $U =I.$ Note that given an orthonomal basis $O_0$ of $\mathbb{H}$ with a fixed $[d_1, \ldots, d_n]$-partition $[Q^{(j)}_0]^n_{j=1}$ such that $Q^{(j)}_0 = \{e^{(j)}_k: 1 \le k \le d_j \}$ ($j=1, \ldots,n$), if $U \in \mathcal{G}^{[Q^{(j)}_0]^n_{j=1}}_{Q_0}$, then for every $Q \in T_U \mathcal{G}^{[Q^{(j)}_0]^n_{j=1}}_{Q_0}$ there exist $n$ complex matrices $X^{(j)} = [ x^{(j)}_{m k}]^{d_j}_{m,k=1}$ ($1 \le j \le n$) such that
\begin{equation}\label{eq:VertVectStrucGroupExpress}
Q = \sum^n_{j=1} \sum^{d_j}_{m,k=1} x^{(j)}_{m k} |e^{(j)}_m \rangle \langle e^{(j)}_k|.
\end{equation}
In particular, each element $Q \in \mathrm{g}^{[Q^{(j)}_0]^n_{j=1}}_{Q_0}$ is of the form
\begin{equation}\label{eq:oAVertVectLieAlg}
Q = \sum^n_{j=1} \sum^{d_j}_{m,k=1} x^{(j)}_{m k} |e^{(j)}_m \rangle \langle e^{(j)}_k|,
\end{equation}
where $X^{(j)} = [ x^{(j)}_{m k}]^{d_j}_{m,k=1}$ is a $d_j \times d_j$ complex matrix for every $j=1,\ldots,n.$ Therefore, $\mathrm{g}^{[Q^{(j)}_0]^n_{j=1}}_{O_0}$ is a linear subspace of $\mathcal{B} (\mathbb{H})$.

The following is the tangent space for the base space $\mathcal{W}^{[Q^{(j)}_0]^n_{j=1}}_{O_0}$ in the operator-theoretic sense.

\begin{definition}\label{df:q-tangvectorBaseSpace}
Fix an orthonomal basis $O_0$ of $\mathbb{H}$ with a given $[d_1, \ldots, d_n]$-partition $[Q^{(j)}_0]^n_{j=1}$.
\begin{enumerate}[{\rm 1)}]

\item A continuous curve $\chi: [a, b] \ni t \mapsto W(t) \in \mathcal{W}^{[Q^{(j)}_0]^n_{j=1}}_{O_0}$ is said to be differential at a fixed $t_0 \in (a, b)$ relative to $[Q^{(j)}_0]^n_{j=1}$, if there is a nonempty subset $\mathcal{A}$ of $\mathcal{B} (\mathbb{H})$ satisfying that for any $Q \in \mathcal{A}$ there exist $\varepsilon>0$ so that $(t_0 -\varepsilon, t_0 + \varepsilon) \subset [a,b]$ and a strongly continuous curve $\gamma: (t_0 -\varepsilon, t_0 + \varepsilon) \ni t \mapsto U_t \in [\Pi^{[Q^{(j)}_0]^n_{j=1}}_{O_0}]^{-1} (O(t))$ such that the limit
$$
\lim_{t \to t_0} \frac{U_t (h) - U_{t_0} (h)}{t - t_0} = Q (h)
$$
for any $h \in \mathbb{H}.$ In this case, $\mathcal{A}$ is called a tangent vector of $\chi$ at $t=t_0$ and denoted by
$$
\mathcal{A} = \frac{d W(t)}{d t}\big |_{t = t_0} = \frac{d \chi (t)}{d t} \big |_{t = t_0}.
$$
We can define the left (or, right) tangent vector of $\chi$ at $t = a$ (or, $t =b$) in the usual way.

\item Given $W \in \mathcal{W}^{[Q^{(j)}_0]^n_{j=1}}_{O_0},$ a tangent vector of $\mathcal{W}^{[Q^{(j)}_0]^n_{j=1}}_{O_0}$ at $W$ relative to $[Q^{(j)}_0]^n_{j=1}$ is define to be a nonempty subset $\mathcal{A}$ of $\mathcal{B} (\mathbb{H}),$ provided $\mathcal{A}$ is a tangent vector of some continuous curve $\chi$ at $t=0,$ where $\chi: (-\varepsilon, \varepsilon) \ni t \mapsto W(t) \in \mathcal{W}^{[Q^{(j)}_0]^n_{j=1}}_{O_0}$ with $\chi (0) = W,$ i.e., $\mathcal{A} = \frac{d W (t)}{d t} \big |_{t=0}.$ We denote by $T_W \mathcal{W}^{[Q^{(j)}_0]^n_{j=1}}_{O_0}$ the set of all tangent vectors of $\mathcal{W}^{[Q^{(j)}_0]^n_{j=1}}_{O_0}$ at $W,$ and write
\be
T \mathcal{W}^{[Q^{(j)}_0]^n_{j=1}}_{O_0} = \bigcup_{W \in \mathcal{W}^{[Q^{(j)}_0]^n_{j=1}}_{O_0}} T_W \mathcal{W}^{[Q^{(j)}_0]^n_{j=1}}_{O_0}.
\ee

\end{enumerate}
\end{definition}

Therefore, the tangent vectors for the base space $\mathcal{W}^{[Q^{(j)}_0]^n_{j=1}}_{O_0}$ are dependent on the orthonomal basis $O_0$ and its $[d_1, \ldots, d_n]$-partition $[Q^{(j)}_0]^n_{j=1}$.

\begin{definition}\label{df:q-tangvectorFiberSpace}
Fix an orthonomal basis $O_0$ of $\mathbb{H}$ with a given $[d_1, \ldots, d_n]$-partition $[Q^{(j)}_0]^n_{j=1}$.
\begin{enumerate}[{\rm 1)}]

\item A strongly continuous curve $\gamma: [a, b] \ni t \mapsto U(t) \in \mathcal{U} (\mathbb{H})$ is said to be differential at a fixed $t_0 \in (a, b),$ if there is an operator $Q \in \mathcal{B} (\mathbb{H})$ such that the limit
$$
\lim_{t \to t_0} \frac{U_t (h) - U_{t_0} (h)}{t - t_0} = Q (h)
$$
for all $h \in \mathbb{H}.$ In this case, $Q$ is called the tangent vector of $\gamma$ at $t=t_0$ and denoted by
$$
Q = \frac{d \gamma (t)}{d t} \Big |_{t = t_0} = \frac{d U (t)}{d t} \Big |_{t = t_0}.
$$
We can define the left (or, right) tangent vector of $\gamma$ at $t = a$ (or, $t =b$) in the usual way.

Moreover, $\gamma$ is called a smooth curve, if $\gamma$ is differential at each point $t \in [a, b],$ and for any $h \in \mathbb{H},$ the $\mathbb{H}$-valued function $t \mapsto \frac{d \gamma (t)}{d t} (h)$ is continuous in $[a, b].$

\item For a given $P \in \mathcal{U} (\mathbb{H}),$ an operator $Q \in \mathcal{B} (\mathbb{H})$ is called a tangent vector of $\xi^{[Q^{(j)}_0]^n_{j=1}}_{O_0}$ at $P,$ if there exists a strongly continuous curve $\gamma: (-\varepsilon, \varepsilon) \ni t \mapsto P_t \in \mathcal{U} (\mathbb{H})$ with $\gamma (0) = P,$ such that $\gamma$ is differential at $t=0$ and $Q = \frac{d \gamma (t)}{d t} \big |_{t =0}.$ Denote by $T_P \xi^{[Q^{(j)}_0]^n_{j=1}}_{O_0}$ the set of all tangent vectors of $\xi^{[Q^{(j)}_0]^n_{j=1}}_{O_0}$ at $P$ relative to $[Q^{(j)}_0]^n_{j=1},$ and write
$$
T \xi^{[Q^{(j)}_0]^n_{j=1}}_{O_0} = \bigcup_{P \in \mathcal{U} (\mathbb{H})} T_P \xi^{[Q^{(j)}_0]^n_{j=1}}_{O_0}.
$$

\item Given $P \in \mathcal{U} (\mathbb{H}),$ a tangent vector $Q \in T_P \xi^{[Q^{(j)}_0]^n_{j=1}}_{O_0}$ is said to be vertical, if there is a strongly continuous curve $\gamma: (-\varepsilon, \varepsilon)\ni t \mapsto P_t \in P \mathcal{G}^{[Q^{(j)}_0]^n_{j=1}}_{Q_0}$ with $\gamma (0) = P$ such that $\gamma$ is differential at $t=0$ and $Q = \frac{d \gamma (t)}{d t} \big |_{t =0}.$ We denote by $V_P \xi^{[Q^{(j)}_0]^n_{j=1}}_{O_0}$ the set of all vertically tangent vectors at $P.$
\end{enumerate}
\end{definition}

\begin{remark}\rm
For any $P \in \mathcal{U} (\mathbb{H}),$ the tangent space $T_P \xi^{[Q^{(j)}_0]^n_{j=1}}_{O_0}$ is the same for any orthonomal basis $O_0$ of $\mathbb{H}$ with a fixed $[d_1, \ldots, d_n]$-partition $[Q^{(j)}_0]^n_{j=1}$, since it is the usual tangent space of $\mathcal{U} (\mathbb{H})$ at $P$ in the operator-theoretic sense. However, the vertically tangent space $V_P \xi^{[Q^{(j)}_0]^n_{j=1}}_{O_0}$ is different from each other for distinct orthonomal bases $O_0$ with a fixed $[d_1, \ldots, d_n]$-partition $[Q^{(j)}_0]^n_{j=1}$. In particular, if $O_0$ has a $[d_1, \ldots, d_n]$-partition $[Q^{(j)}_0]^n_{j=1}$ such that $ Q^{(j)}_0 = \{e^{(j)}_k: 1 \le k \le d_j \}$ ($j=1, \ldots,n$), then every $Q \in V_P \xi^{[Q^{(j)}_0]^n_{j=1}}_{O_0}$ has the form
\begin{equation}\label{eq:VertTangentVect}
Q = \sum^n_{j=1} \sum^{d_j}_{m,k=1} x^{(j)}_{m k} P |e^{(j)}_m \rangle \langle e^{(j)}_k|,
\end{equation}
where $X^{(j)} = [ x^{(j)}_{m k}]^{d_j}_{m,k=1}$ is a $d_j \times d_j$ complex matrix for every $j=1,\ldots,n.$
\end{remark}

Given $G \in \mathcal{G}^{[Q^{(j)}_0]^n_{j=1}}_{O_0},$ the right action $R_G$ of $\mathcal{G}^{[Q^{(j)}_0]^n_{j=1}}_{O_0}$ on $\xi^{[Q^{(j)}_0]^n_{j=1}}_{O_0}$ defined by
$$
R_G (U) = U G,\quad \forall U \in \mathcal{U} (\mathbb{H}),
$$
induces a map $(R_G)_*: T_P \xi^{[Q^{(j)}_0]^n_{j=1}}_{O_0} \mapsto T_{R_G(P)} \xi^{[Q^{(j)}_0]^n_{j=1}}_{O_0}$ for each $P \in \mathcal{U} (\mathbb{H})$ such that
$$
(R_G)_* (Q) = Q G,\quad \forall Q \in T_P \xi^{[Q^{(j)}_0]^n_{j=1}}_{O_0}.
$$
Since $R_G$ preserves the fibers of $\xi^{[Q^{(j)}_0]^n_{j=1}}_{O_0},$ then $(R_G)_*$ maps $V_P \xi^{[Q^{(j)}_0]^n_{j=1}}_{O_0}$ into $V_{R_G(P)} \xi^{[Q^{(j)}_0]^n_{j=1}}_{O_0}.$

Now, we are ready to define the concept of quantum connection over the operator-principal fiber bundle $\xi^{[Q^{(j)}_0]^n_{j=1}}_{O_0}.$

\begin{definition}\label{df:q-connetion}
Fix an orthonomal basis $O_0$ of $\mathbb{H}$ with a given $[d_1, \ldots, d_n]$-partition $[Q^{(j)}_0]^n_{j=1}$. A non-Abelian quantum connection $\Omega^{[Q^{(j)}_0]^n_{j=1}}$ on the operator-principal fiber bundle \be
\xi^{[Q^{(j)}_0]^n_{j=1}}_{Q_0} = (\mathcal{U} (\mathbb{H}), \Pi^{[Q^{(j)}_0]^n_{j=1}}_{Q_0}, \mathcal{W}^{[Q^{(j)}_0]^n_{j=1}}_{Q_0})
\ee
is a family of linear functionals
\be
\Omega^{[Q^{(j)}_0]^n_{j=1}} = \{\Omega^{[Q^{(j)}_0]^n_{j=1}}_P:\; P \in \mathcal{U} (\mathbb{H}) \},
\ee
where for each $P \in \mathcal{U} (\mathbb{H}),$ $\Omega^{[Q^{(j)}_0]^n_{j=1}}_P$ is a linear mapping defined in $T_P \xi^{[Q^{(j)}_0]^n_{j=1}}_{Q_0}$ with values in $\mathrm{g}_{O_0}^{[Q^{(j)}_0]^n_{j=1}},$ satisfying the following conditions:
\begin{enumerate}[{\rm (1)}]

\item For any $P \in \mathcal{U} (\mathbb{H})$ and for all vertically tangent vectors $Q \in V_P \xi^{[Q^{(j)}_0]^n_{j=1}}_{Q_0},$ one has
\begin{equation}\label{eq:q-ConnectionVertTangVect}
\Omega_P (Q) = P^{-1} Q.
\end{equation}

\item $\Omega^{[Q^{(j)}_0]^n_{j=1}}_P$ depends continuously on $P,$ in the sense that if $P_n$ converges to $P$ as well as $Q_n \in T_{P_n} \xi^{[Q^{(j)}_0]^n_{j=1}}_{Q_0}$ converges $Q_0 \in T_P \xi^{[Q^{(j)}_0]^n_{j=1}}_{Q_0}$ in the operator topology of $\mathcal{B} (\mathbb{H}),$ then
\be
\lim_{n \to \infty} \Omega^{[Q^{(j)}_0]^n_{j=1}}_{P_n} (Q_n) = \Omega^{[Q^{(j)}_0]^n_{j=1}}_P (Q_0)
\ee
in $\mathrm{g}^{[Q^{(j)}_0]^n_{j=1}}_{O_0}.$

\item Under the right action of $\mathcal{G}^{[Q^{(j)}_0]^n_{j=1}}_{Q_0}$ on $\xi^{[Q^{(j)}_0]^n_{j=1}}_{O_0},$ $\Omega^{[Q^{(j)}_0]^n_{j=1}}$ transforms according to
\begin{equation}\label{eq:GaugeTransConnection}
\Omega^{[Q^{(j)}_0]^n_{j=1}}_{R_G(P)} [(R_G)_* (Q )] = G^{-1} \Omega^{[Q^{(j)}_0]^n_{j=1}}_P (Q) G,
\end{equation}
for $G \in \mathcal{G}^{[Q^{(j)}_0]^n_{j=1}}_{Q_0},$ $P \in \mathcal{U} (\mathbb{H}),$ and $Q \in T_P \xi^{[Q^{(j)}_0]^n_{j=1}}_{O_0}.$

\end{enumerate}
Such a connection is simply called an $[Q^{(j)}_0]^n_{j=1}$-connection.
\end{definition}

Next, we present a canonical example of non-Abelian quantum connections, which plays a crucial role in the expression of the non-Abelian observable-geometric phases.

\begin{example}\label{Ex:CanonicalConnection}\rm
Fix an orthonomal basis $O_0$ of $\mathbb{H}$ with a given $[d_1, \ldots, d_n]$-partition $[Q^{(j)}_0]^n_{j=1}$ such that $Q^{(j)}_0 = \{e^{(j)}_k: 1 \le k \le d_j \}$ ($1 \le j \le n$). We define $\check{\Omega}^{[Q^{(j)}_0]^n_{j=1}} = \{\check{\Omega}^{[Q^{(j)}_0]^n_{j=1}}_P: P \in \mathcal{U} (\mathbb{H}) \}$ as follows: For each $P \in \mathcal{U} (\mathbb{H}),$ $\check{\Omega}^{[Q^{(j)}_0]^n_{j=1}}_P : T_P \xi^{[Q^{(j)}_0]^n_{j=1}}_{O_0} \mapsto \mathrm{g}^{[Q^{(j)}_0]^n_{j=1}}_{O_0}$ is defined by
\begin{equation}\label{eq:CanonConnction}
\check{\Omega}^{[Q^{(j)}_0]^n_{j=1}}_P (Q) = P^{-1} \star_{[Q^{(j)}_0]^n_{j=1}} Q
\end{equation}
for any $Q \in T_P \xi^{[Q^{(j)}_0]^n_{j=1}}_{O_0},$ where
$$
P^{-1} \star_{[Q^{(j)}_0]^n_{j=1}} Q = \sum^n_{j = 1} \sum^{d_j}_{m,k=1} \langle e^{(j)}_m, P^{-1} Q e^{(j)}_k \rangle |e^{(j)}_m \rangle \langle e^{(j)}_k|.
$$

Indeed, by \eqref{eq:VertTangentVect} one has
\be
P^{-1} \star_{[Q^{(j)}_0]^n_{j=1}} Q = P^{-1} Q \in \mathrm{g}^{[Q^{(j)}_0]^n_{j=1}}_{O_0}
\ee
for any $Q \in V_P \xi^{[Q^{(j)}_0]^n_{j=1}}_{O_0},$ namely $\check{\Omega}^{[Q^{(j)}_0]^n_{j=1}}_P$ satisfies \eqref{eq:q-ConnectionVertTangVect}. The conditions (2) and (3) of Definition \ref{df:q-connetion} are clearly satisfied by $\check{\Omega}^{[Q^{(j)}_0]^n_{j=1}}.$ Hence, $\check{\Omega}^{[Q^{(j)}_0]^n_{j=1}}$ is an $[Q^{(j)}_0]^n_{j=1}$-connection on $\xi^{[Q^{(j)}_0]^n_{j=1}}_{O_0}.$ In this case, we write $\check{\Omega}^{[Q^{(j)}_0]^n_{j=1}}_P = P^{-1} \star_{[Q^{(j)}_0]^n_{j=1}} d P$ for any $P \in \mathcal{U} (\mathbb{H}).$
\end{example}

\subsection{Quantum parallel transportation}\label{q-ParallelTransport}

The quantum parallel transportation in the state space was introduced in \cite{Simon1983, AA1987} and studied in \cite{Anandan1992} in details. In \cite{Chen2020}, the quantum parallel transportation over the observable space was studied. Next, we continue to study non-Abelian quantum parallel transport.

\begin{definition}\label{df:q-lift}
Fix an orthonomal basis $O_0$ of $\mathbb{H}$ with a given $[d_1, \ldots, d_n]$-partition $[Q^{(j)}_0]^n_{j=1}$. For a continuous curve $\mathbf{C}_W: [a, b] \ni t \longmapsto W(t) \in \mathcal{W}^{[Q^{(j)}_0]^n_{j=1}}_{O_0},$ a lift of $\mathbf{C}_W$ with respect to $[Q^{(j)}_0]^n_{j=1}$ is defined to be a continuous curve
$$
\mathbf{C}_P: [a, b] \ni t \longmapsto U(t) \in \mathcal{U} (\mathbb{H})
$$
satisfying that for each $t\in [a, b],$ $U(t) \in (\Pi^{[Q^{(j)}_0]^n_{j=1}}_{O_0})^{-1}[W(t)]$.

Such a lift $\mathbf{C}_P$ is called a $[Q^{(j)}_0]^n_{j=1}$-lift of $\mathbf{C}_W.$
\end{definition}


\begin{definition}\label{df:SmoothCurve}
Fix an orthonomal basis $O_0$ of $\mathbb{H}$ with a given $[d_1, \ldots, d_n]$-partition $[Q^{(j)}_0]^n_{j=1}$. A continuous curve $\mathbf{C}_W: [a, b] \ni t \longmapsto W(t) \in \mathcal{W}^{[Q^{(j)}_0]^n_{j=1}}_{O_0}$ is said to be smooth relative to $[Q^{(j)}_0]^n_{j=1}$, if it has a $[Q^{(j)}_0]^n_{j=1}$-lift $\mathbf{C}_P: [a, b] \ni t \longmapsto U(t) \in \mathcal{U} (\mathbb{H})$ which is a smooth curve. In this case, $\mathbf{C}_P$ is called a smooth $[Q^{(j)}_0]^n_{j=1}$-lift of $\mathbf{C}_W.$
\end{definition}

Note that, if a continuous curve $\mathbf{C}_W: [a, b] \ni t \longmapsto W(t) \in \mathcal{W}^{[Q^{(j)}_0]^n_{j=1}}_{O_0}$ is smooth, then it is differential at every point $t \in [a, b].$ Indeed, suppose that $\mathbf{C}_P: [a, b] \ni t \longmapsto U(t) \in \mathcal{U} (\mathbb{H})$ is a smooth $[Q^{(j)}_0]^n_{j=1}$-lift of $\mathbf{C}_W.$ For each $t \in [a, b],$ we have $\frac{d \mathbf{C}_P (t)}{d t} \in \frac{d W(t)}{d t},$ namely $\frac{d W(t)}{d t}$ is a nonempty subset of $\mathcal{B} (\mathbb{H}),$ and hence $\mathbf{C}_W$ is differential.

Now, we are ready to define the parallel transportation with respect to a non-Abelian quantum connection as follows.

\begin{definition}\label{df:q-HorizontalLift}
Fix an orthonomal basis $O_0$ of $\mathbb{H}$ with a given $[d_1, \ldots, d_n]$-partition $[Q^{(j)}_0]^n_{j=1}$. Let $\Omega^{[Q^{(j)}_0]^n_{j=1}}$ be a $[Q^{(j)}_0]^n_{j=1}$-connection on $\xi^{[Q^{(j)}_0]^n_{j=1}}_{O_0}.$ Suppose that $\mathbf{C}_W: [0, T] \ni t \longmapsto W(t) \in \mathcal{W}^{[Q^{(j)}_0]^n_{j=1}}_{O_0}$ be a smooth curve relative to $[Q^{(j)}_0]^n_{j=1}$. If $\mathbf{C}_P: [0, T] \ni t \longmapsto U(t) \in \mathcal{U} (\mathbb{H})$ is a smooth $[Q^{(j)}_0]^n_{j=1}$-lift of $\mathbf{C}_W$ such that
\begin{equation}\label{eq:ParallelTransfConnectionCond}
\Omega^{[Q^{(j)}_0]^n_{j=1}}_{U(t)} \Big [ \frac{d U(t)}{d t}\Big ] =0
\end{equation}
for every $t \in [0, T],$ then $\mathbf{C}_P$ is called a horizontal lift of $\mathbf{C}_W$ with respect to $\Omega^{[Q^{(j)}_0]^n_{j=1}}.$

In this case, $\mathbf{C}_P$ is simply called the horizontal $[Q^{(j)}_0]^n_{j=1}$-lift of $\mathbf{C}_W$. And, the curve $\mathbf{C}_P: t \mapsto U(t)$ is called the parallel transportation along $\mathbf{C}_W$ with the starting point $\mathbf{C}_P (0) = U(0)$ with respect to the connection $\Omega^{[Q^{(j)}_0]^n_{j=1}}$ on $\xi^{[Q^{(j)}_0]^n_{j=1}}_{O_0}.$
\end{definition}

The following proposition shows the existence of the horizontal lifts.

\begin{proposition}\label{prop:ParallelTransf}\rm
Fix an orthonomal basis $O_0$ of $\mathbb{H}$ with a given $[d_1, \ldots, d_n]$-partition $[Q^{(j)}_0]^n_{j=1}$. Let $\Omega^{[Q^{(j)}_0]^n_{j=1}}$ be a $[Q^{(j)}_0]^n_{j=1}$-connection on $\xi^{[Q^{(j)}_0]^n_{j=1}}_{O_0}.$ Suppose that $\mathbf{C}_W: [0, T] \ni t \longmapsto W(t) \in \mathcal{W}^{[Q^{(j)}_0]^n_{j=1}}_{O_0}$ is a smooth curve relative to $[Q^{(j)}_0]^n_{j=1}$. For any $U_0 \in (\Pi^{[Q^{(j)}_0]^n_{j=1}}_{O_0})^{-1} [W(0)],$ there exists a unique horizontal $[Q^{(j)}_0]^n_{j=1}$-lift $\mathbf{C}_P$ of $\mathbf{C}_W$ with the initial point $\mathbf{C}_P (0) = U_0.$
\end{proposition}

\begin{proof}
Let $\Gamma: [0, T] \ni t \longmapsto U (t) \in \mathcal{U} (\mathbb{H})$ be a smooth $[Q^{(j)}_0]^n_{j=1}$-lift of $\mathbf{C}_W$ with $\Gamma (0) = U_0.$ To prove the existence, note that the condition (2) of Definition \ref{df:q-connetion} implies that the function $t \mapsto \Omega^{[Q^{(j)}_0]^n_{j=1}}_{\Gamma (t)} \big [ \frac{d \Gamma (t)}{d t} \big ]\in \mathrm{g}_{O_0}^{[Q^{(j)}_0]^n_{j=1}}$ is continuous  in $[0, T].$ Then,
\begin{equation}\label{eq:GeodesicEquaGaugeTransf}
\frac{d G (t)}{d t} = - \Omega^{[Q^{(j)}_0]^n_{j=1}}_{\Gamma (t)} \Big [ \frac{d \Gamma (t)}{d t} \Big ] \cdot G (t)
\end{equation}
with $G (0) = I$ has the unique solution in $\mathrm{g}_{O_0}^{[Q^{(j)}_0]^n_{j=1}}$ in $[0, T].$ By computation, we know that $\mathbf{C}_P (t) = \Gamma (t) \cdot G (t)$ is the required horizontal $[Q^{(j)}_0]^n_{j=1}$-lift of $\mathbf{C}_W$ for the initial point $U_0 \in (\Pi^{[Q^{(j)}_0]^n_{j=1}}_{O_0})^{-1} [W(0)].$

To prove the uniqueness, suppose $\mathbf{\check{C}}_P: [0, T] \ni t \longmapsto \check{U} (t) \in\mathcal{U} (\mathbb{H})$ be another horizontal $[Q^{(j)}_0]^n_{j=1}$-lift of $\mathbf{C}_W$ for the initial point $U_0 \in (\Pi^{[Q^{(j)}_0]^n_{j=1}}_{O_0})^{-1} [W(0)].$ Then $\mathbf{\check{C}}_P (t) = \mathbf{C}_P (t) \cdot \check{G} (t)$ for all $t \in [0, T],$ where $\check{G} (t) \in \mathrm{g}_{O_0}^{[Q^{(j)}_0]^n_{j=1}}$ such that $\check{G}(0) = I.$ Since
$$
0 = \Omega_{\check{U}(t)} \Big [ \frac{d \check{U}(t)}{d t} \Big ] = \check{G}(t)^{-1} \frac{d \check{G}(t)}{d t},
$$
this follows that $\check{G}(t) = I$ for all $t \in [0, T].$ Hence, the horizontal $[Q^{(j)}_0]^n_{j=1}$-lift of $\mathbf{C}_W$ is unique for the initial point $U_0 \in (\Pi^{[Q^{(j)}_0]^n_{j=1}}_{O_0})^{-1} [W(0)].$
\end{proof}

\begin{example}\label{Ex:nAqParallelTransport}\rm
Fix an orthonomal basis $O_0$ of $\mathbb{H}$ with a given $[d_1, \ldots, d_n]$-partition $[Q^{(j)}_0]^n_{j=1}$ such that $Q^{(j)}_0 = \{e^{(j)}_k: 1 \le k \le d_j \}$ ($1 \le j \le n$). Suppose that $\mathbf{C}_P: [0,T] \ni t \mapsto U(t) \in \mathcal{U} (\mathbb{H})$ is a unitary evolution satisfying the operator Schr\"{o}dinger equation
\begin{equation}\label{eq:SchrodingerEquTimeUnitaryEvolution}
\mathrm{i} \frac{d U(t)}{d t} = H(t) U (t)
\end{equation}
where $H(t)$'s are time-dependent Hamiltonian operators in $\mathbb{H}.$ We define $\mathbf{C}_W: [0, T] \ni t \longmapsto W(t) \in \mathcal{W}^{[Q^{(j)}_0]^n_{j=1}}_{O_0}$ by $W(t) = \Pi^{[Q^{(j)}_0]^n_{j=1}}_{O_0} [U(t)]$ for all $t \in [0, T]$. In what follows, we construct a horizontal $[Q^{(j)}_0]^n_{j=1}$-lift of $\mathbf{C}_W$ with respect to the non-Abelian quantum connection $\check{\Omega}^{[Q^{(j)}_0]^n_{j=1}}$, which is the canonical $[Q^{(j)}_0]^n_{j=1}$-connection introduced in Example \ref{Ex:CanonicalConnection}.

For $1 \le j \le n,$ let $\tilde{V}_j(t) = [\tilde{v}^{(j)}_{m k} (t)]^{d_j}_{m,k=1}$ be the solution to the matrix equation
\beq\label{eq:GeoPhaseMatrix}
\mathrm{i} \frac{d}{d t} \tilde{V}_j (t) = C_j (t) \tilde{V}_j (t),
\eeq
with $\tilde{V}_j(0) = I_{n_i}$, where $C_j (t) = [c^{(j)}_{m k} (t)]^{d_j}_{m,k=1}$ with $c^{(j)}_{m,k} (t) = - \langle  U(t) e^{(j)}_m, H (t)  U(t) e^{(j)}_k \rangle .$ All $\tilde{V}_j (t)$'s are $d_j \times d_j$ unitary matrices, since $C_j (t)$'s are all Hermitian matrices. Define $\mathbf{\tilde{C}}_P: [0,T] \ni t \mapsto \tilde{U}(t) \in \mathcal{U} (\mathbb{H})$ by
$$
\tilde{U} (t) = \sum^n_{j=1} \sum^{d_j}_{m,k=1} \tilde{v}^{(j)}_{m k} (t) U(t) | e^{(j)}_m \rangle \langle e^{(j)}_k|
$$
for every $t \in [0, T],$ along with the initial point $\tilde{U}(0) = U(0) \in (\Pi^{[Q^{(j)}_0]^n_{j=1}}_{O_0})^{-1} [W(0)].$ Then $\mathbf{\tilde{C}}_P$ is a smooth $[Q^{(j)}_0]^n_{j=1}$-lift of $\mathbf{C}_W$ such that
$$
\check{\Omega}^{[Q^{(j)}_0]^n_{j=1}}_{\tilde{U} (t)} \Big [ \frac{d \tilde{U}(t)}{d t} \Big ] =0
$$
for all $t \in [0, T],$ Thus, $\mathbf{\tilde{C}}_P$ is the horizontal $[Q^{(j)}_0]^n_{j=1}$-lift of $\mathbf{C}_W$ with respect to $\check{\Omega}^{[Q^{(j)}_0]^n_{j=1}},$ namely $\mathbf{\tilde{C}}_p$ is the parallel transportation along $\mathbf{C}_W$ with the starting point $\mathbf{C}_P (0) = U(0)$ with respect to the connection $\check{\Omega}^{[Q^{(j)}_0]^n_{j=1}}$ on $\xi^{[Q^{(j)}_0]^n_{j=1}}_{O_0}.$
\end{example}

\subsection{Quantum holonomy}

Fix an orthonomal basis $O_0$ of $\mathbb{H}$ with a given $[d_1, \ldots, d_n]$-partition $[Q^{(j)}_0]^n_{j=1}$. Let $\Omega^{[Q^{(j)}_0]^n_{j=1}}$ be a $[Q^{(j)}_0]^n_{j=1}$-connection on $\xi^{[Q^{(j)}_0]^n_{j=1}}_{O_0}.$ Suppose that $\mathbf{C}_W: [0, T] \ni t \longmapsto W(t) \in \mathcal{W}^{[Q^{(j)}_0]^n_{j=1}}_{O_0}$ is a smooth closed curve relative to $[Q^{(j)}_0]^n_{j=1}$. For any $U_0 \in (\Pi^{[Q^{(j)}_0]^n_{j=1}}_{O_0})^{-1} [W(0)],$ if there exists a horizontal $[Q^{(j)}_0]^n_{j=1}$-lift $\mathbf{C}_P$ of $\mathbf{C}_W$ with the initial point $\mathbf{C}_P (0) = U_0,$ then $\mathbf{C}_P (T) \in (\Pi^{[Q^{(j)}_0]^n_{j=1}}_{O_0})^{-1} [W(0)]$ and hence there exists $G \in \mathrm{g}_{O_0}^{[Q^{(j)}_0]^n_{j=1}}$ such that $\mathbf{C}_P (T) = U_0 G$. In this case, $G$ is called a {\it quantum holonomy element} associated with $U_0$, the connection $\Omega^{[Q^{(j)}_0]^n_{j=1}}$, and the curve $\mathbf{C}_W$. If we choose different closed curve $\mathbf{C}_W$, we will obtain different group elements $G$. These group elements depends on the non-Abelian quantum connection $\Omega^{[Q^{(j)}_0]^n_{j=1}}$ on the operator-principal fiber bundle $\xi^{[Q^{(j)}_0]^n_{j=1}}_{O_0}$ and the initial point $U_0 =\mathbf{C}_P (0)$ of the horizontal lift. The set of such elements which correspond to a point $U_0 \in \mathcal{U} (\mathbb{H})$ form a subgroup of the structure group $\mathrm{g}_{O_0}^{[Q^{(j)}_0]^n_{j=1}}$ of $\xi^{[Q^{(j)}_0]^n_{j=1}}_{O_0}$. This subgroup is called the {\it quantum holonomy group} of $\Omega^{[Q^{(j)}_0]^n_{j=1}}$ associated with $U_0$.

\subsection{Geometric interpretation of non-Abelian observable-geometric phases}\label{GeoInter}

We are now ready to give a geometric interpretation of $G_j$'s defined in \eqref{eq:nAOGP} in Section \ref{NOGP}. To this end, using the notations involved in Section \ref{NOGP}, we first fix an orthonormal basis $O_0 = \{\psi^{(j)}_k: k=1,\ldots, d_j;\; 1\le j \le n\}$ of $\mathbb{H}$ with a given $[d_1, \ldots, d_n]$-partition $[Q^{(j)}_0]^n_{j=1}$ such that $Q^{(j)}_0 = \{\psi^{(j)}_k: 1 \le k \le d_j \}$ ($1 \le j \le n$). Then, we define $\tilde{U} (t) \in \mathcal{U} (\mathbb{H})$ for $0 \le t \le T$ by
$$
\tilde{U} (t) = \sum^n_{j = 1} \sum^{d_j}_{k=1} |\tilde{\psi}^{(j)}_k (t) \rangle \langle \psi^{(j)}_k |,
$$
where $\tilde{\psi}^{(j)}_k (t)$'s are defined in \eqref{eq:ParallCurveState}. Let
\be
\bar{U}(t) = \sum^n_{j = 1} \sum^{d_j}_{k=1} |\bar{\psi}^{(j)}_k (t) \rangle \langle \psi^{(j)}_k |,
\ee
where $\bar{\psi}^{(j)}_k (t)$'s are defined as in Proposition \ref{prop:nAOGP}. Note that $\bar{U}(T) = \bar{U}(0)$ by the construction of $\bar{\psi}^{(j)}_k (t)$'s. Define $\mathbf{C}_W: [0, T] \ni t \mapsto W(t) = \Pi^{[Q^{(j)}_0]^n_{j=1}}_{O_0} [\bar{U}(t)]$, which is a smooth closed curve in $\mathcal{W}^{[Q^{(j)}_0]^n_{j=1}}_{O_0}.$ Thus,
$$
\mathbf{\tilde{C}}_P: \; [0, T] \ni t \longmapsto \tilde{U} (t) \in \mathcal{U} (\mathbb{H})
$$
is a smooth $[Q^{(j)}_0]^n_{j=1}$-lift of $\mathbf{C}_W,$ since $\tilde{U} (t) = \bar{U} (t) \bar{G} (t),$ where
\be
\bar{G}(t) = \sum^n_{j = 1} \sum^{d_j}_{m,k=1} \bar{v}^{(j)}_{m k} (t) |\psi^{(j)}_m \rangle \langle \psi^{(j)}_k | \in \mathrm{g}_{O_0}^{[Q^{(j)}_0]^n_{j=1}},
\ee
for every $0 \le t \le T.$

\begin{proposition}\label{prop:GeoInterNaOGP}\rm
Using the above notions, $\mathbf{\tilde{C}}_P$ is the horizontal $[Q^{(j)}_0]^n_{j=1}$-lift of $\mathbf{C}_W$ with respect to $\check{\Omega}^{[Q^{(j)}_0]^n_{j=1}}$ in the operator-principal fiber bundle $\xi^{[Q^{(j)}_0]^n_{j=1}}_{O_0}$, where $\check{\Omega}^{[Q^{(j)}_0]^n_{j=1}}$ is the $[Q^{(j)}_0]^n_{j=1}$-connection defined in Example \ref{Ex:CanonicalConnection}. Therefore, $\mathbf{\tilde{C}}_P: \;[0, T] \ni t \mapsto \tilde{U} (t)$ is the parallel transportation along $\mathbf{C}_W$ with respect to the connection $\check{\Omega}^{[Q^{(j)}_0]^n_{j=1}}$ on $\xi^{[Q^{(j)}_0]^n_{j=1}}_{O_0}.$
\end{proposition}

\begin{proof}
By \eqref{eq:ParallCondState}, we have for $1 \le j \le n,$
\be
\langle \psi^{(j)}_m, \tilde{U}^{-1}(t) \frac{d}{d t} \tilde{U} (t) \psi^{(j)}_k \rangle = \langle \tilde{\psi}^{(j)}_m (t), \frac{d}{d t} \tilde{\psi}^{(j)}_k (t)\rangle =0,\quad \forall 1\le m,k \le d_j,
\ee
and hence,
\begin{equation}\label{eq:CanonicalParallelCond}
\check{\Omega}^{[Q^{(j)}_0]^n_{j=1}}_{\tilde{U} (t)} \Big [ \frac{d \tilde{U} (t)}{d t}  \Big ] = 0,\quad \forall t \in [0,T].
\end{equation}
This concludes the required result.
\end{proof}

\begin{remark}\label{rk:GeoInterNaOGP}\rm
Since
\be
\tilde{U} (T) = \sum^n_{j=1} \sum^{d_j}_{m,k=1} g^{(j)}_{m k} |\psi^{(j)}_m\rangle \langle \psi^{(j)}_k|,
\ee
then the non-Abelian observable-geometric phases $G_j = [g^{(j)}_{m k}]^{d_j}_{m,k=1}$ ($1 \le j \le n$) defines a holonomy element $\tilde{U}(T) \in \mathrm{g}_{O_0}^{[Q^{(j)}_0]^n_{j=1}}$ of $\check{\Omega}^{[Q^{(j)}_0]^n_{j=1}}$ associated with the initial point $\mathbf{\tilde{C}}_P (0) = I$ and the curve $\mathbf{C}_W$. Note that, by Proposition \ref{prop:nAOGP}, $G_j$ depends only on the flow of Hilbert subspaces $\mathbf{K}_j: [0,T] \ni t \mapsto \mathbb{H}_j (t)$ for every $1 \le j \le n$. Thus, the holonomy element $\tilde{U}(T)$ does not depend on the curve $\mathbf{C}_W$, but depends only on the orthonormal basis $O_0 = \{\psi^{(j)}_k: k=1,\ldots, d_j;\; 1\le j \le n\}$ of $\mathbb{H}$ with the $[d_1, \ldots, d_n]$-partition $[Q^{(j)}_0]^n_{j=1}$ and $n$ Hilbert subspace flows $[\mathbf{K}_j]^n_j$. This explains the geometrical meaning of $\tilde{U}(T)$ in terms of the non-Abelian geometric phases $[G_j]^n_{j=1}$.
\end{remark}

\


\bibliography{apssamp}

\end{document}